\documentstyle[12pt]{article}
\newcommand{\be}{\begin{equation}}
\newcommand{\ee}{\end{equation}}
\newcommand{\ba}{\begin{eqnarray}}
\newcommand{\ea}{\end{eqnarray}}
\begin{document}
\rightline{FUSAV-96/05}
\rightline{MAY 1996$~~~~$}
\vspace*{1cm}
\begin{center}
\renewcommand{\thefootnote}{\dag}
{\large {\bf
HIGH ENERGY
DENSITY FLUCTUATIONS IN TERMS OF 
FACTORIAL MOMENTS AND \\

\smallskip
ASSOCIATED FREQUENCY MOMENTS}}\footnote{To be published in 
Int'l J. Mod. Phys. A  }
\end{center}

\smallskip
 ~\\
 ~\\
\renewcommand{\thefootnote}{\alph{footnote}}
\begin{center}
\setcounter{footnote}{0}
Mikul\' a\v s Bla\v zek\footnote{Electronic address: 
blazek@savba.savba.sk} \\
 ~\\
{\em 
Institute of Physics, Slovak Academy of Sciences, \\
842 28  Bratislava, Slovak Republic  }
\end{center}

\vskip1truecm
\begin{center}
\begin{quotation}
 ~~  ~~~~~~ ~~~~~~~~~~~   ~~~ ~~~ ~~~  ABSTRACT \\

\medskip
We propose to verify relations between quantities which characterize
scaling properties of high energy density fluctuations in terms of
factorial moments and newly introduced associated frequency moments.
Typical examples are presented in frame of systematics developed
in the present paper. It involves also several sorts of moments 
applied so far in 
the search for intermittency and multifractality. Our approach takes
the advantage of relations in which a linear combination of 
associated 
frequency moments reduces statistical fluctuations to the same extent
as it is done by corresponding factorial moments.
\end{quotation}
\end{center}

\bigskip

\medskip
\noindent
\noindent
{\bf 1. Introduction} 

\smallskip
It can be taken already as firmly established that several high energy 
particle density fluctuations reveal multifractality${{}^{1,2}}$. This 
phenomenon appears quite clearly in lepton, hadron as well as heavy 
ion 
induced collisions and it is observed at different energies of 
colliding
objects.

High energy intermittency, or in general, multifractality is usually 
described in terms of quantities
characterizing scaling properties of factorial${^{3,4}}$, $F(q)$, and
frequency${^{5,6,7}}$, $G(q)$, moments. From this point of view, the
corresponding scaling indices ($a(q)$  and $\tau(q)$, respectively)
represent the most frequent quantities one is dealing with in this 
region
of physics.
 
It is usually argued that in the case of very large 
multiplicities\footnote{It will be seen later that, in general, the 
term
 ``multiplicity'' mentioned at this place, means a quantity
which depends on the form of individual moments under consideration.},
 both
sets of scaling indices are related quite simply, namely,
\be
a(q) = q-1-\tau(q) ~~.
\ee
Data available so far show a statistically significant difference 
between 
the left and right hand sides of (1), compare Table 1. This 
circumstance
is not surprising because a consistent description of realistic 
experimental data available so far should be represented by a relation
extending (1) which takes into account finite multiplicities.

In the present contribution that extention is given in explicit form, 
for
different definitions of the factorial and associated frequency 
moments; they
are related by the requirement that at very large 
multiplicities\footnote{See the preceding 
footnote.} the following type of relation should be satisfied,
\be
F(q) = \omega(q) G(q)~~,
\ee
$\omega(q)$ being present due to the appropriate normalization.  It 
will be
seen also that for a given finite value of the order $q$ the number 
of 
correction terms involved in that extension, is finite.
 
A short and incomplete survey of the factorial moments (represented 
by  the
ratio ``numerator over denumerator'') which are (or might be) 
currently 
applied in high energy physics is seen in  Table 2.

Introductory relations mentioned in the next Section give the 
possibility
 to include (into
the relations between scaling indices) the correction terms appearing
due to the finite value of the corresponding multiplicities. In the 
third
Section we discuss first of all the standard approaches while 
the next  Section deals with several other  forms of the factorial and 
associated frequency moments. We emphasize also (in the fifth Section)
 that
there is missed a systematic search  for multifractality in high 
energy
physics; so far, in individual cases, there are not known reliable
conditions
which allow to specify the most appropriate type of statistical
moments for the study of multifractality. Conclusions and suggestions
are formulated in the last Section.

\bigskip
\noindent
{\bf 2. Preliminaries} 

\smallskip

We consider charged particle density distributions in a 
(pseudo)rapidity 
window $\Delta Y$; let $n_{me}$ denote  the number of particles 
observed 
in the $m$-th bin ($m$=1,2,...,$M$) of the $e$-th event 
($e$=1,2,...,$E$),
the equal size of bins being $\delta y = \Delta Y/M$. The 
non-normalized and 
non-averaged ``primitive'' factorial moment is represented by quantity
$F(n_{me};q)$, say, with $q \geq 1$,
\ba
F(n;q)  &=& n(n-1)...(n-q+1)   \nonumber       \\
        &=& n^q + A_1 n^{q-1} + ... + A_{q-1}n   \nonumber  \\
        &=& n^q\left [ 1 + \frac{A_1(q)}{n} + ...+ 
                           \frac{A_{q-1}(q)}{n^{q-1}} \right ]
\ea
where (compare${^{10}}$),
\ba
 A_1 &=& A_1(q) = -q(q-1)/2 ~, \nonumber  \\
 A_2 &=& A_2(q) = +q(q-1)(q-2)(3q-1)/24 ~,  \nonumber  \\
 A_3 &=& A_3(q) = -q^2 (q-1)^2 (q-2)(q-3)/48 ~,
\ea
etc. and with vanishing $A_j(q)$ if $q \leq j$, j = 1,2,... .

In the present paper the assumption is applied that scaling properties
identified by several appropriate  statistical moments are observed 
in  a
range of the number of bins, $M$, say,  
\begin{center}
$  M_{low} \le M \le M_{high}~~$; 
\end{center}
$M_{high}$ is usually given by resolution of the devices applied. In 
high
energy physics that quantity is finite (like e.g. also in turbulence
where it is known as the Kolmogorov' bound).

We take also as granted that the slopes and intercepts entering the 
leading
order asymptotic terms which characterize the scaling properties of 
the
moments under consideration, are sufficiently accurate so that the 
next to the
leading order terms are much smaller (in absolute value) than the 
former
ones.

Now we show the extension of eq.(1) in several concrete cases.

\bigskip
\noindent
{\bf 3. Standard approaches} 

\smallskip
{\em Standard approaches} involve following form of factorial
and associated frequency moments:

\smallskip
\noindent
{\bf 3.1. Horizontal moments} 

If properties of individual events are to be
emphasized it is usually expected that the {\em horizontal} moments
are more appropriate for investigating the scaling properties of a
sample of events. In this respect e.g. the muon-proton and
muon-deuteron collisions${^7}$ at 280 GeV  
as well as the 800 GeV protons interacting with emulsion nuclei${^8}$ 
can be mentioned.

The standard horizontal factorial moments, $F_e^{(H)}(q)$, 
characterizing the $e$-th event
are introduced
by\footnote{To minimize the degree of possible confusion in notation
and naming of different moments we write down explicitly all indices
involved in respective quantities and summations.}
(compare with the case [X(A/C)] of Table 2; there
are quoted also other references),
\be
 F_e^{(H)} (q) = M^{q-1} \sum_{m=1}^M F(n_{me};q)/[N_e^{(H)}]^q
\ee
and their vertical averaging gives the full form,
\be
F^{(H)}(q) =  \frac{1}{E} \sum_{e=1}^E F_e^{(H)}(q) ~~.
\ee
Denominator of the horizontal moment (5), $N_e^{(H)}$,
\be
 N_e^{(H)} = \sum_{m=1}^M n_{me}
\ee
represents the multiplicity of the $e$-th event.

The well known (horizontal) frequency $G$-moments characterizing the
$e$-th event,
\be
G_e^{(H)} (q) =  \left \{ \sum_{m=1}^M n_{me}^q / [\sum_{m=1}^M 
n_{me}]^
q \right \} \vartheta (n_{me} - q)
\ee
after vertical averaging lead to
\be
G^{(H)}(q) =\frac{1}{E} \sum_{e=1}^E G_e^{(H)}(q)~~.
\ee
   
Frequency moments (8) represent the leading order asymptotic term 
of (5)
(up to the normalization) in the limit of very large number of
observed particles (a more accurate specification is given below).

Now, we are looking for a relation between the factorial and 
associated
frequency moments just introduced.

Since the low values of multiplicity $n_{me}, ~~n_{me} <q$, do not
influence the former moments (due to their definition, eq.(5)), the 
presence
of $\vartheta$-function in (8) [with $\vartheta (x) = 0$ if $x<0$
and $\vartheta (x) = 1$ if $x \geq 0$] guarantees the same property
of the later ones, eq.(8).

\smallskip
Let us adopt the view  that fractal structure is revealed by factorial 
moments if the following scaling property is observed,
\be
\frac{1}{E} \sum_{e=1}^E F_e^{(H)} (q)~~ \propto~~ f_q^{(H)} 
M^{a_q^{(H)}}~~.
\ee 
And the fractal structure is detected by frequency moments if the 
following
scaling property is satisfied,
\be
\frac{1}{E} \sum_{e=1}^E [N_e^{(H)}]^{{}^{- \lambda}} G_e^{(H)} (q)
~~\propto~~ [N_0^{(G),(H)}]^{{}^{-\lambda}} g_q^{(H)} M^{-\tau_q^
{(H)}} ~.
\ee

In eq.(11), $N_0^{(G),(H)}$ represents the {\em effective average
multiplicity}; $N_e^{(H)}$ is given by (7), and $\lambda$ = 0,1,2,...;
with $\lambda$ = 0 the usual form of the scaling property is obtained
(as it is applied in${^{5,6}}$  and in many other papers). With 
respect to
eq.(3),
\be
F_e^{(H)}(q) =M^{q-1} \{ G_e^{(H)}(q) + \frac{A_1 (q)}{N_e^{(H)}}
G_e^{(H)}(q-1) + \cdots +\frac{A_{q-1} (q)} 
  {[N_e^{(H)}]^{{}^{q-1}}} G_e^
{(H)} (1) \}
\ee
and we conclude that the asymptotic equality (of the form (2)) between 
moments  characterizing a given event,
\be
 F_e^{(H)} (q) = M^{q-1} G_e^{(H)} (q) ~,
\ee
is satisfied  as far as the (total) multiplicity of that event, 
$N_e^{(H)}$,
is very large,
\be
 N_e^{(H)} \gg 1 ~~.
\ee

The series expansion on r.h.s. of (12) retains finite number of 
terms;
this means that such a linear combination  of the frequency 
$G$-moments 
reduces statistical fluctuations (contained in all individual 
frequency 
moments) to the same extent as it is done by the corresponding 
factorial 
moment appearing on the l.h.s. of (12).

Averaging eq.(12) over events,
\ba
\frac{1}{E} \sum_{e=1}^E F_e^{(H)} (q) &=& M^{q-1} 
\left \{ \frac{1}{E}  \sum
_{e=1}^E G_e^{(H)} (q)+ \frac{A_1}{E} \sum_{e=1}^E [N_e^{(H)}]^{-1}
G_e^{(H)}(q-1) + \cdots \right . \nonumber  \\
    &+&\left. \frac{A_{q-1}}{E} \sum_{e=1}^E [N_e^{(H)}]^
{{}^{-q+1}} G_e^{(H)} (1) \right \}
\ea
and application of the asymptotics (10) and (11) leads to 
the result\footnote{We 
assume that both asymptotics, (10) and (11), are satisfied 
simultaneously; similar note is applied also in other cases considered
   in 
the present contribution.},
\ba
f_q^{(H)} M^{a_q^{(H)} -q +1} &=& g_q^{(H)} M^{-\tau_q^{(H)}} +
\frac{A_1}{N_0^{(G),(H)}} g_{q-1} M^{-\tau_{q-
1}^{(H)}} + \cdots    \nonumber  \\
    &+&\frac{A_{q-1}}{[N_0^{(G),(H)}]^{{}^{q-1}}} g_1^{(H)} 
M^{-\tau_1^{(H)}}.
\ea
Equation (16) allows to conclude that the asymptotics (1) is 
satisfied as far as
\be
N_0^{(G),(H)}~~\gg ~~1~.
\ee
Moreover, with $q=2$  eq.(16) 
gives\footnote{In all cases considered in the present paper,
factorial as well as associated frequency moments with $q=1$ are equal
to unity. With respect to that circumstance the corresponding slopes 
($a_1$
and $\tau_1$) identically vanish while the intercepts ($f_1$ and 
$g_1$) are
equal to unity. However, with respect to a better transparency of the
corresponding relations, those quantities are not replaced by their 
numerical
value.},
\be
f_2^{(H)} M^{a_2^{(H)} - 1} = g_2^{(H)} M^{-\tau_2^{(H)}} -
\frac{g_1^{(H)}}{N_0^{(G),(H)}} M^{-\tau_1^{(H)}}
\ee
(where $A_1(2) = -1$ is applied) and with $q=3$ eq.(16) and (18) 
lead to 
\be
f_3^{(H)} M^{a_3^{(H)} -2} + \frac{2f_2^{(H)}}{N_0^{(G),(H)}} 
M^{a_2^{(H)}-1}
= g_3^{(H)} M^{-\tau_3^{(H)}} - \frac{g_2^{(H)}}{N_0^{(G),(H)}} 
M^{-\tau_2^
{(H)}}~.
\ee
As far as experimental data allow to extract
the value of quantities
\be
\{ f_2^{(H)}, a_2^{(H)}; g_2^{(H)}, \tau_2^{(H)}; N_0^{(G),(H)} \}~,
\ee
eq.(18) allows to verify a consistency relation between them. Then
eq.(19) represents a relation between $\{ f_3^{(H)},a_3^{(H)};
g_3^{(H)}, \tau_3^{(H)} \}$  and the quantities quoted in (20).
Analogous conclusion can be formulated\footnote{The same comment
refers also to other cases considered in the present paper.}
if the value $q=4,5,...$ is
inserted in (16).

Quantities characterizing scaling properties of standard horizontal 
moments are related by (16); we propose to apply that relation  
or its special cases (like (18), (19) and other ones obtained with 
$q=4,5,...$) 
for verification of the consistency or for prediction of some of 
them\footnote{See the preceding footnote.}.
  
Let us note that eq.(12) allows to express an associated frequency 
moment 
$G_e(q)$ as a linear combination of the factorial moments $F_e(q)$
with decreasing order $q$, namely,  \\
$$ G_e^{(H)}(q)
= \sum_{j=1}^q \left \{ B(q-j+1)/[N_e^{(H)}]^{q-j+1} \right \} 
F_e^{(H)}(q-
j+1) ~; $$
\medskip
\noindent
the coefficients $B(q-j+1)$ being expressed in terms of the
coefficients $A_j(q)$, compare${}^{10}$. Now, let us average 
the aforementioned relation  over
events and apply the asymptotics $(1/E)\sum_{e=1}^EG_e(q)~ \propto~ g_
q M^{-\tau_q}$ together 
with
\be 
 \frac{1}{E} \sum_{e=1}^E [N_e^{(H)}]^
{-\lambda} 
F_e^{(H)}(q)~ \propto~
[N_0^{(F),(H)}]^{-\lambda} f_q^{(H)} M^{a_q^{(H)}} ~. 
\ee
Comparison of the result which is obtained by this procedure, with 
eq.(12)
leads to the following interesting conclusion,

\medskip
\be 
 N_0^{(F),(H)}~ =~ N_0^{(G),(H)}~ \equiv ~ N_0^{(H)} 
\ee
where $N_0^{(G),(H)}$ is introduced by (11); this means that the 
effective
average multiplicity characterizing horizontal factorial moments is 
the same as that one characterizing  associated frequency moments 
(this
conclusion is valid also in other cases treated below).

\bigskip
\noindent
{\bf 3.2. Vertical moments} 

\smallskip
If rare events with sharp peaks are not to be missed${^5}$, the 
{\em vertical} analysis is suggested. In this case the normalized
standard vertical factorial moments characterizing a given (say, the
$m$-th) bin acquire the form\footnote{The factor $M^{q-1}$ guarantees
 that
eq.(1) is satisfied in the corresponding asymptotics.} (compare
[Y(A/C)] of Table 2),
\be
F_m^{(V)}(q) = M^{q-1} E^{q-1} \sum_{e=1}^E 
  F(n_{me};q)/[N_m^{(V)}]^{{}^q}
\ee
and the consecutive horizontal averaging leads to the full form,
\be
F^{(V)}(q) = \frac{1}{M} \sum_{m=1}^M F_m^{(V)}(q)
\ee
with
\be
N_m^{(V)} = \sum_{e=1}^E n_{me}
\ee
being the sum of multiplicities which appear in the $m$-th bin of all
 events.
Factorial moments of the form (23), up to the normalization factor,
are successfully applied e.g. when there are investigated interactions
of 200 A GeV sulphur nuclei with gold target${^{11}}$, as well as in 
several
other cases (compare e.g. Table 2).
The associated vertical frequency moments,
\be
G_m^{(V)}(q) = \left \{ \sum_{e=1}^E n_{me}^q /[
N_m^{(V)}]^{{}^q} \right \} \vartheta( n_{me} - q)
\ee
averaged horizontally, give the full moments
\be
G^{(V)}(q) =
\frac{1}{M} \sum_{m=1}^M G_m^{(V)}(q)~.
\ee
Now, the fractal structure is observed in terms of the factorial
 moments if
the scaling relation
\be
 \frac{1}{M} \sum_{m=1}^M F_m^{(V)}(q)~~\propto ~~ E^{q-1} 
f_q^{(V)} M^
{a_q^{(V)}}
\ee
is satisfied, and in terms of the associated frequency moments, if 
it holds,
\be
\frac{1}{M} \sum_{m=1}^M [N_m^{(V)}]^{{}^{-\lambda}} G_m^{(V)}(q)~~
\propto~~ [N_0^{(G),(V)}]^{{}^{-\lambda}} g_q^{(V)} 
M^{-\tau_q^{(V)}}~.
\ee
Application of (3) allows to arrive at the relation,
\ba
~~~~~ ~~~~ ~~~~ ~~F_m^{(V)}(q) &=& M^{q-1} E^{q-1} 
\left \{ G_m^{(V)}(q) +
\frac{A_1}{N_m^{(V)}} G_m^{(V)}(q-1) + 
\cdots  ~~~~ ~~~~ ~~~~ ~~~~ ~~ \right .  \nonumber  \\  
     &+&\left. \frac{A_{q-1}}{[N_m^{(V)}]^{{}^{q-1}}} G_m^
{(V)}(1) \right \}~. ~ 
\ea
It is seen that a simple relation of the form (2) between moments 
characterizing the $m$-th bin,
\begin{center}
 $F_m^{(V)}(q) ~=~ M^{q-1} E^{q-1} G_m^{(V)}(q) $
\end{center}
is satisfied if the multiplicity $N_m^{(V)}$, eq.(25), is sufficiently
large, i.e.,
\be
N_m^{(V)}~~\gg~~1~.
\ee
Averaging (30) over bins and application of the asymptotics (28) and
(29) gives,
\ba
f_q^{(V)} M^{a_q^{(V)} -q +1}~&=&~ g_q^{(V)} M^{-\tau_q^{(V)}} +
\frac{A_1}{N_0^{(G),(V)}} g_{q-1}^{(V)} M^{-\tau_{q-1}^{(V)}} +
\cdots  \nonumber  \\
&+&\frac{A_{q-1}}{[N_0^{(G),(V)}]^{{}^{q-1}}} g_1^{(V)} 
M^{-\tau_1^{(V)}}~.
\ea
Now, the asymptotic equation (1) is recovered as far as the effective
average multiplicity $N_0^{(G),(V)}$ is sufficiently large, i.e.,
\be
 N_0^{(G),(V)}~~\gg~~1~.
\ee

With $q=2$ and $q=3$, we obtain from (32), respectively,
\be
f_2^{(V)} M^{a_2^{(V)} - 1}~=~ g_2^{(V)} M^{-\tau_2^{(V)}} -
 \frac{g_1^{(V)}}{N_0^{(G),(V)}} M^{-\tau_1^{(V)}}
\ee 
and applying (34),
\be
f_3^{(V)} M^{a_3^{(V)} -2} + \frac{2}{N_0^{(G),(V)}} f_2^{(V)} M^{a_2^
{(V)} -1} = g_3^{(V)} M^{-\tau_3^{(V)}} - \frac{g_2^{(V)}}
{N_0^{(G),(V)}} M^{-\tau_2^{(V)}}~.
\ee
Equations (34),(35) and other ones obtained with $q=4,5,...$ are
analogous to the preceding case (compare with (18) and (19)).
Let us add that also in this case, if the effective average 
multiplicity
$N_0^{(F),(V)}$ characterizing the vertical factorial moments, is
introduced in an analogous way as in the preceding part (compare with
eq.(21)) the following equality is obtained, $N_0^{(F),(V)} =
N_0^{(G),(V)} \equiv N_0^{(V)}$.

\bigskip
\noindent
{\bf 3.3. Mixed moments} 

\smallskip
Besides the horizontal and vertical analyses also a mixed
approach is applied.

\smallskip
{\bf (i)} Especially, factorial moments of the form (compare with
the case [Z(A/C)] in Table 2),
\ba
F^{(HV)}(q) ~&=&~ \frac{1}{M} \sum_{m=1}^M \frac{\frac{1}{E} 
\sum_{e=1}^
 E F(n_{me};q)} {\left[ \frac{1}{M} \frac{1}{E} \sum_{e=1}^E N_e^{
(H)} \right ]^{{}^q}}  \nonumber  \\
 &=& M^{q-1} E^{q-1}\frac{\sum_{m=1}^M \sum_{e=1}^E F(n_{me};q)}{
 [N^{(HV)}]^{{}^q} }
\ea
are applied for instance in${^{12}}$ when intermittency in muon-proton
scattering at 280 GeV is investigated. The associated G-function is 
now
expressed in the following form,
\be
G^{(HV)}(q) ~=~ \frac{\sum_{m=1}^M \sum_{e=1}^E n_{me}^q}{
[N^{(HV)}]^{{}^q}}~ \vartheta (n_{me} - q)~,
\ee
$N^{(HV)}$ in (36) and (37) being the total number of charged 
particles
observed in the sample of $E$-events,
\be
 N^{(HV)}~=~ \sum_{m=1}^M \sum_{e=1}^E n_{me}
\ee
and the number of all charged particles in the $e$-th event, 
$N_e^{(H)}$,
entering (36) is given by (7).

If the corresponding fractal structure is present in the data, 
validity
of the following scaling properties is expected,
\be
F^{(HV)}(q) ~~\propto~~ E^{q-1} f_q^{(HV)} M^{a_q^{(HV)}}~,
\ee
and
\be
G^{(HV)}(q)~~\propto~~ g_q^{(HV)} M^{-\tau_q^{(HV)}}~.
\ee
In this case, application of (3) leads to the following relation
between factorial (36) and associated frequency (37) moments,
\ba
F^{(HV)}(q)~&=&~ M^{q-1} E^{q-1} \left \{ G^{(HV)}(q) +
 \frac{A_1}{N^{(HV)}} G^{(HV)}(q-1) + \cdots \right .  \nonumber  \\
& & \left. 
+~ \frac{A_{q-1}}{[N^{(HV)}]^{{}^{q-1}}} G^{(HV)}(1) \right \}~.
\ea
In the asymptotics (39) and (40), eq.(41) gives,
\be
f_q^{(HV)} M^{a_q^{(HV)} -q+1} ~=~ X_q^{(HV)}
\ee
where
\ba
 X_q^{(HV)} ~\equiv ~ g_q^{(HV)} M^{-\tau_q^{(HV)}} +
\frac{A_1}{N^{(HV)}} g_{q-1}^{(HV)} M^{-\tau_{q-1}^{(HV)}} 
+ \cdots  \nonumber   \\
+ \frac{A_{q-1}}{[N^{(HV)}]^{{}^{q-1}}} g_1^{(HV)} 
M^{-\tau_1^{(HV)}}~.
\ea
Now, eq.(41) is reduced to (2) and eq.(42) to (1) as far as the
multiplicity $N^{(HV)}$, rel.(38), satisfies the condition
\be
 N^{(HV)} ~~\gg~~ 1~.
\ee
If $q=2$ we obtain from (42),
\be
f_2^{(HV)} M^{a_q^{(HV)}-1}~=~ g_2^{(HV)} M^{-\tau_2^{(HV)}} -
\frac{g_1^{(HV)}}{N^{(HV)}} M^{-\tau_1^{(HV)}} ~,
\ee
and with $q=3,4,...$ the relations follow which are similar to the
preceding cases.

\medskip
{\bf (ii)} To {\em reduce} the influence of non-flat rapidity 
distribution varying within the finite bin width $\delta y = 
\Delta Y/M$, introduction of the following factor was suggested
in${^{13}}$, 
\be
R(q)~=~\frac{ \frac{1}{M} \sum_{m=1}^M \langle \rho_{me} \rangle ^q}{
\left[ \frac{1}{M} \sum_{m=1}^M \langle \rho_{me} \rangle \right 
]^q}~~;
\ee
angular brackets denote averaging over all events and $\rho _{me} =
n_{me}/\delta y$ means particle density in the $m$-th bin of the 
$e$-th event,
so that $\langle \rho _{me} \rangle $ is the value of single particle
rapidity distribution. In our notation, the factor (46) can be 
expressed
as it follows,
\be
R(q) ~=~ M^{q-1} \frac{\sum_{m=1}^M [N_m^{(V)}]^{{}^q}}{
[N^{(HV)}]^{{}^q}}~.
\ee
Now, the reduced factorial moment $F^{(red')}(q)$,
$$ F^{(red')}(q)~=~\frac{F^{(HV)}(q)}{R(q)} $$
(numerator being expressed by (36) and denominator by (47))
acquires the form (compare with ${\rm [Z(A/D_1)]}$ of Table 2),
\be
F^{(red')}(q)~=~ 
E^{q-1} \frac{\sum_{m=1}^M \sum_{e=1}^E F(n_{me};q)}{
\sum_{m=1}^M [N_m^{(V)}]^{{}^q}}
\ee
and the multiplicity $N_m^{(V)}$ is given by (25).  At this point, 
let us introduce\footnote{See the preceding footnote.} the
factorial moment $F^{(red)}(q)$,
\be 
F^{(red)}(q)~=~ M^{q-1} F^{(red')}(q)~.
\ee

The frequency moments (37) are associated also with (49) (as well as
with (48)),
\be
G^{(red)}(q)~=~\frac{ \sum_{m=1}^M \sum_{e=1}^E n_{me}^q}{
\sum_{m=1}^M[N_m^{(V)}]^{{}^q}}~ \vartheta (n_{me} - q)~.
\ee
As it is seen, there is a remarkable difference between the couple
(36) + (37) and (49) + (50).

Now, the scaling properties are formulated in the form,
\ba
F^{(red)}(q) ~~ &\propto& ~~ E^{q-1} f_q^{(red)} M^{a_q^{(red)}}~, \\
G^{(red)}(q) ~~&\propto& ~~ g_q^{(red)} M^{-\tau_q^{(red)}}~.
\ea

Applying an analogous procedure as in the preceding case, the relation 
between
 factorial and associated frequency moments reads,
\ba
F^{(red)}(q)~=~ E^{q-1} \left \{ G^{(red)}(q) +A_1 \frac{\sum_{m=1}^M 
[N_m^
{(V)}]^{{}^{q-1}}}{\sum_{m=1}^M [N_m^{(V)}]^{{}^q}} G^{(red)}(q-1)
+ \cdots  \right .  \nonumber   \\
+ \left.  A_{q-1} \frac{\sum_{m=1}^M N_
m^{(V)}}{\sum_{m=1}^M [N_m^{(V)}]^{{}^q}} G^{(red)}
(1) \right \}  ~~.~~~~~ ~~~~~~~~~~    
\ea
In this case the factorial moment $F^{(red)}(q)$, (49), is reduced
(up  to the normalization) to the associated frequency moment 
$G^{(red)}(q)$,
(50), if the multiplicity $N_m^{(V)}$, (25), satisfies the
following relation,
\be
 \frac {  \sum_{m=1}^M [N_m^{(V)}]^{{}^{q-j}} } { 
\sum_{m=1}^M [N_m^{(V)}]^
{{}^q} }~~ \ll~~1~,
\ee
with $j=1,2,...,q-1$. The condition (54) differs considerably 
from (44). 

In the  asymptotics (51) and (52), eq.(53) gives,
\be
f_q^{(red)} M^{a_q^{(red)} -q+1} ~=~ X_q^{(red)}~,
\ee
where
\ba
X_q^{(red)}~\equiv~ g_q^{(red)} M^{-\tau_q^{(red)}} + A_1 \frac{
\sum_{m=1}^M [N_m^{(V)}]^{{}^{q-1}}} {\sum_{m-1}^M [N_m^{(V)}]^{{}^q}}
g_{q-1}^{(red)} M^{-\tau_{q-1}^{(red)}} + \cdots   \nonumber  \\
+ A_{q-1} \frac{\sum_{m-1}^M N_m^{(V)}}{\sum_{m=1}^M 
[N_m^{(V)}]^{{}^q}}
g_1^{(red) } M^{-\tau_1^{(red)}}   ~.~~   ~~~~
\ea
With $q=2$, eq.(56) leads to,
\be
f_2^{(red)} M^{a_2^{(red)} -1} ~=~ g_2^{(red)} M^{-\tau_2^{(red)}} -
\frac{\sum_{m=1}^M N_m^{(V)} }{\sum_{m=1}^M [N_m^{(V)}]^{{}^2}} 
g_1^{(red)}
 M^{-\tau_1^{(red)}}  ~.
\ee
If the condition (54) is satisfied then (55) with (56) induce
relation (1) between scaling indices $a_q^{(red)}$ and 
$\tau_q^{(red)}$.

\bigskip  
\noindent
{\bf 4. Some non-standard cases} 

\smallskip
To demonstrate the existence of somehow {\em more involved situations}
we mention at least two following cases:

\medskip  
\noindent
{\bf 4.1. Collisions of lead nuclei with nuclear target}

\smallskip
Study of 158 GeV/n lead collisions with nuclear target published
in${^{14}}$ applies an
interesting form of the factorial moment proposed in${^{15}}$ (compare
with the case ${\rm [Z(A/G_1)]}$ of Table 2),
\be
F^{(x1)}(q)~=~ M^{q-1} E^{q-1} \frac{ \sum_{m=1}^M 
\sum_{e=1}^E F(n_{me};
q)}{ \sum_{m=1}^M F(N_m^{(V)};q) } ~~.
\ee

With respect to (3), where the function $F(n_{me};q)$ is introduced,
the moment (58) can be expressed as it follows,
\be
F^{(x1)}(q)~=~ \frac{ M^{q-1} E^{q-1} }{V^{(M)}(q)}~  
\frac{\sum_{m=1}^M \sum_
{e=1}^E F(n_{me};q)} { \sum_{m=1}^M [N_m^{(V))}]^{{}^q}}
\ee
or, with respect to (49),
\be
 F^{(x1)}(q)~=~ \frac{1}{V^{(M)}(q)} F^{(red)}(q)~.
\ee
In both expresions just mentioned,
\be
V^{(M)}(q)~ \equiv~ 1 + A_1 \frac{\sum_{m=1}^M 
[N_m^{(V)}]^{{}^{q-1}} }
{\sum_{m=1}^M [N_m^{(V)}]^{{}^q}} + \cdots + A_{q-1} \frac{ 
\sum_{m=1}^M
N_m^{(V)}} {\sum_{m=1}^M [N_m^{(V)}]^{{}^q}}~.
\ee

The frequency moments $G^{(x1)}(q)$ associated with (60) (as well 
as e.g.
in the case ${\rm [Z(A/D_1)]}$ of Table 2) acquire the form of eq.(50),
especially\footnote{It is not surprising that different forms of 
factorial
moments lead to the same  form of the leading order asymptotic term.},
\be
G^{(x1)}(q)~=~ G^{(red)}(q)~.
\ee
Now, by means of (59) and (62) we obtain,
\ba
F^{(x1)}(q)~=~\frac{   M^{q-1} E^{q-1} }{ V^{(M)}(q) } \left \{ 
G^{(x1)}(q) 
+ A_1 \frac{ \sum_{m=1}^M [N_m^{(V)}]^{{}^{q-1}}}{\sum_{m=1}^M 
[N_m^{(V)}]^{{}^q}} G^{(x1)}(q-1) + \cdots \right .   \nonumber   \\
  \left. + A_{q-1} \frac{\sum_{m=1}^M N_m^{(V)}} {\sum_{m=1}^M 
[N_m^{(V)}]^{{}^q}} G^{(x1)}(1) \right \}~. ~~~~ ~~~~ ~~~~ ~~~~ ~~~
 ~~~~~~~~
\ea
The appropriate scaling properties are formulated in the following 
way\footnote{The fact that factor $V^{(M)}(q)$ entering eq.(60)
depends on the number of bins, $M$, prevents us to identify, in
general, the intercepts $f_q^{(x1)}$ with $f_q^{(red)}$ and the
slopes $a_q^{(x1)}$ with $a_q^{(red)}$, as they are introduced by
(64) and (51), respectively.},
\ba
F^{(x1)}(q)~~&\propto&~~E^{q-1} f_q^{(x1)} M^{a_q^{(x1)}}~,  \\
G^{(x1)}(q) ~~&\propto&~~ g_q^{(x1)} M^{-\tau_q^{(x1)}}
\ea
(bearing in mind eq.(59)).

In this case, eq.(63) allows to obtain
\be
f_q^{(x1)} M^{a_q^{(x1)}-q+1} ~=~ \frac{1}{V^{(M)}(q)}~  
X^{(red)}(q)~,
\ee
$X^{(red)}(q) \equiv X_q^{(red)}$  being given by (56). If the 
multiplicities $N_m^{(V)}$ are such that condition (54) is valid
then (a) eq.(66) reveals that the scaling indices introduced
by (64) and (65) satisfy rel.(1), and (b)  eq.(63) leads to the
asymptotic relation (2). 
Applying a similar procedure like in preceeding cases,   
the following relation between lowest order scaling characteristics
is obtained from (66) with $q=2$,
\be
f_2^{(x1)} M^{a_2^{(x1)} -1}~=~ \frac{g_2^{(x1)} M^{-\tau_2^{(x1)}} -
\frac{\sum_{m=1}^M N_m^{(V)}}{\sum_{m=1}^M [N_m^{(V)}]^{{}^2}}~ 
g_1^{(x1)}
M^{-\tau_1^{(x1)}} }
{ 1 - \frac{\sum_{m=1}^M N_m^{(V)}}{\sum_{m=1}^M 
[N_m^{(V)}]^{{}^2}}  }~.
\ee

\medskip  
\noindent
{\bf 4.2. Carbon-cooper interactions} 

The factorial moments\footnote{See footnote${}^i$.} suggested 
essentially
in${^{15}}$ and applied in the intermittency analysis of the 
carbon-cooper collisions at 4.5 A GeV/c in${^{16}}$ (compare 
with ${\rm [Y(A/E)]}$
of Table 2),
\be
F_m^{(x2)}(q)~=~ M^{q-1} E^{q-1} \frac{ \sum_{e=1}^E F(n_{me};q)}
{F(N_m^{(V)};q)}
\ee
can be expressed in the form
\be
F_m^{(x2)}(q)~=~ \frac{1}{V(N_m^{(V)};q)}~F_m^{(V)}(q)
\ee
where the factorial moments $F_m^{(V)}(q)$ characterize the standard
 vertical 
case, (23), and
\ba
V(N_m^{(V)};q)~&=&~ 1+ \frac{A_1}{N_m^{(V)}}+\cdots+ 
\frac{A_{q-1}}{[N_m^
{(V)}]^{{}^{q-1}}} \nonumber   \\
 &\equiv& \frac{F(N_m^{(V)};q)}{[N_m^{(V)}]^{{}^q}} ~,
\ea
the bin multiplicity $N_m^{(V)}$ is given by (25). 

A slight adaptation of (69) and averaging over bins lead to the 
relation,
\be
\frac{1}{M} \sum_{m=1}^M V(N_m^{(V)};q) F_m^{(x2)}(q)~=~
\frac{1}{M} \sum_{m=1}^MF_m^{(V)}(q)~.
\ee
In the present case, let us expect scaling property of the factorial 
moments in an extended form, namely,
\be
\frac{1}{M} \sum_{m=1}^M [N_m^{(V)}]^{{}^{-\lambda}} F_m^{(x2)}(q)
~\propto~ E^{q-1} [N_0^{(F)}]^{{}^{-\lambda}} f_q^{(x2)} 
M^{a_q^{(x2)}}
\ee
where $N_0^{(F)}$ is the effective average multiplicity characterizing 
the factorial moments under consideration (and $\lambda =0,1,2,...$). 
With respect to the scaling 
property (28)  of the moments $F_m^{(V)}(q)$, eq.(68) allows to 
conclude\footnote{Compare with observation of${^{17}}$ saying that the
choice of a different normalization factor does not change the 
dependence
of factorial moments on the number of bins, $M$, as far as this factor 
is
proportional to $M^q$.},
\be 
a_q^{(x2)}~=~ a_q^{(V)}
\ee
and
\be
f_q^{(x2)} \left \{ 1 + \frac{A_1}{N_0^{(F)}} + \cdots + 
\frac{A_{q-1}}
{[N_0^{(F)}]^{{}^{q-1}}}  \right \} ~~=~~ f_q^{(V)} ~.
\ee	
Frequency moments $G_m^{(x2)}(q)$ associated with the factorial moments
(68), satisfy the relation,
\be
G_m^{(x2)}(q)~=~ \frac{\sum_{e=1}^E n_{me}^q}{[N_m^{(V)}]^{{}^q}}~
\vartheta
(n_{me}-q)~=~ G_m^{(V)}(q)~,
\ee
the last quantity characterizing the standard vertical case, compare 
eq.(26). Scaling properties of those frequency moments are assumed 
also
in an extended form,
\be
\frac{1}{M} \sum_{m=1}^M [N_m^{(V)}]^{{}^{-\lambda}} G_m^{(x2)}(q)~~
\propto~~ [N_0^{(G)}]^{{}^{-\lambda}} g_q^{(x2)} M^{-\tau_q^{(x2)}}
\ee
where the effective average multiplicity $N_0^{(G)}$ specifies scaling
properties of the frequency moments under consideration (and $\lambda
=0,1,2,...,$ as it is usual in this paper).

\smallskip
Relation between moments introduced by (68) and (75) acquires the form,
\be
F_m^{(x2)}(q)~=~ \frac{M^{q-1} E^{q-1}}{V(N_m^{(V)};q)}~Y_m^{(x2)}(q)
\ee
with
\be
Y_m^{(x2)}~\equiv~ G_m^{(x2)}(q) + \frac{A_1}{N_m^{(V)}} G_m^{(x2)}(
q-1) + \cdots + \frac{A_{q-1}}{[N_m^{(V)}]^{{}^{q-1}}} G_m^{(x2)}(1)~.
\ee

Let us express (77) in two alternative forms, namely,
\be
V(N_m^{(V)};q)~F_m^{(x2)}(q)~=~ (ME)^{q-1} ~ Y_m^{(x2)}(q)
\ee
and
\be
F_m^{(x2)}(q)~=~(ME)^{q-1}~[V(N_m^{(V)};q)]^{{}^{-1}}~Y_m^{(x2)}(q)
\ee
where
\be
[V(N_m^{(V)};q)]^{{}^{-1}} ~=~c_0 + \frac{c_1}{N_m^{(V)}}+\frac{c_2}{
[N_m^{(V)}]^{{}^2}}+ \cdots
\ee
with $c_0=1, c_1 = -A_1, c_2=A_1^2 - A_2 , \ldots$, and $A_j$ 
given by (4).
In this case,
averaging over bins and application of the scaling properties (72) and
(76) in  (79) and (80) allow to achieve equality between the effective
average multiplicities,
\be
N_0^{(F)}~=~ N_0^{(G)}~\equiv~ N_0^{(x2)}~.
\ee
Therefore, we obtain from (77),
\ba
f_q^{(x2)} M^{a_q^{(x2)} -q+1} &=&\frac{1}{V(N_0^{(x2)};q)}~\left 
\{ g_q^{(x2)}
 M^{-\tau_q^{(x2)}} + \frac{A_1}{N_0^{(x2)}}g_{q-1}^{(x2)} M^{-
\tau_q^{(x2)}} + \cdots \right.  \nonumber  \\
&+&\left. \frac{A_{q-1}}{[N_0^{(x2)}]^{{}^{q-1}}} g_1^{(x2)}M^
{-\tau_1^{(
x2)}} \right \} ,
\ea
$V(N;q)$ being given by (70). Especially, with $q=2$,
\be
f_2^{(x2)} M^{a_2^{(x2)} - 1} ~=~ \frac{g_2^{(x2)} M^{-\tau_2^{(x2)}}-
\frac{1}{N_0^{(x2)}}  g_1^{(x2)} M^{-\tau_1^{(x2)}}}{1-\frac{1}
{N_0^{(x2)}}} ~.
\ee
In this case equation (1) between scaling indices is recovered by (83)
as far as the effective average multiplicity $N_0^{(x2)}$ is 
sufficiently
large, i.e.,
\be
N_0^{(x2)}~\gg~1~
\ee
while relation of the form (2) between the factorial $F_m^{(x2)}(q)$, 
eq.(68), and the associate frequency $G_m^{(x2)}(q)$, eq.(75), 
moments is obtained if the multiplicity $N_m^{(V)}$ is very large,
\be
N_m^{(V)}~ \gg~1~,
\ee
as it is also in the standard vertical case, compare eq.(31).

\bigskip  
\noindent
{\bf 5. Short survey}

\smallskip
We summarize shortly the main items of the present contribution:

Search for intermittency in high energy data on density fluctuations
allows to recognize three sorts of factorial moments which are
currently applied (compare columns $X, Y, Z,$ in Table 2  and the
corresponding references e.g. in line $C$ where some successfull
forms can be found).

Concrete formes of factorial moments discussed in the present 
paper are
considered as representative samples of the types included into 
Table 2,
this Table being not complete, several other expressions can be added 
there. Every form of the factorial moment $F(q)$ can be accompanied by
an associateed frequency moment $G(q)$. There exists a relation between
(i) the factorial moment $F(q)$ and a linear combination of finite 
number
of terms involving the frequency moments $G(q)$ with decreasing order
$q$; and (ii) the frequency $G(q)$ moment and a linear combination of 
finite
number of terms involving the factorial moments $F(q)$ of decreasing
order $q$.

It is well known${^{3,4}}$, that convenient averaging reduces the 
influence
of statistical fluctuations in factorial moments. Therefore, the linear 
combination of frequency moments just mentioned in case (i) reduces
 that 
influence of noise to the same extent as it is done by the 
corresponding
factorial moment. And, on the other hand, a linear combination of the
factorial moments mentioned in case (ii) gives rise to the noise of the
same level as it is done by the corresponding frequency moment.

Associated frequency moments represent asymptotics of the corresponding
factorial moments; in the present contribution, in every case, 
there are
clearly specified the multiplicities which allow to achieve that 
asymptotics.
Let us note that the relation  mentioned above in item (i), can be
interpreted also in the following way: a finite number of
asymptotic (=frequency) moments allows to reconstruct the full
(=corresponding factorial) moment. Hence, that relation represents
a solution of the appropriate (moments) inverse problem expressed in
the form ``How to restore 
the full form of a statistical moment by means of its asymptotics ?'' . 

Scaling properties of the statistical moments involved are described in 
terms of slopes (scaling indices), intercepts and if convenient also
effective average multiplicities; all those quantities are 
experimentally
accessible. In every case discussed in the present contribution,
the aforementioned relation
between factorial and linear combination of associated frequency 
moments is 
presented also in terms of quantities which characterize those scaling
properties. The form of relation which is obtained in this way allows
to recover eq.(1) between scaling indices, namely, in the asymptotics
also specified in the present contribution.   In turn, that relation 
might serve either as predicting some quantities or as consistency 
conditions for data to be considered as sufficiently accurate ones.

While the slopes $a(q)$ characterize to a high extent dynamical
fluctuations, the slopes $\tau(q)$ can be considered as solutions of
 extended fundamental equation${^{18,19}}$ of the Halsey 
et al. type\footnote{
That equation can be considered also as a definition of the scaling
indices $\tau = \tau(q)$ for arbitrary real value of the order $q$.}.
This circumstance also points out the importance of a link between 
those 
two sets of slopes.

\bigskip  
\noindent
{\bf 6. Conclusions} 

\smallskip
In the present contribution we formulate (exact) relations between 
quantities which characterize scaling properties of factorial and 
associated frequency moments. 
No additional experiment nor any new measurement is needed for 
verifying
those relations by existing data; all information necessary  for that
purpose is involved in the data presently available\footnote{Only the 
data 
on scaling indices are published usually.}.

Systematic analyses of the present as well as the near future data
might help to classify the sort of fluctuations\footnote{It is 
understood 
here that quantities characterizing those fluctuations satisfy 
relations 
advocated in the present contribution. Moreover, the fractal phenomena 
usually reveal their presence only if they are analysed by appropriate
quantities; negative results  were published, too, concerning e.g.
$S+Au$~~ Ref.${^{20}}$, and $O+S$~ ion-emulsion${^{21}}$ interactions 
at 200 GeV/n.},
as well as their dependence on the type of collision, energy, kinematic
cuts, etc.

In this region of physics, sufficiently reliable Monte Carlo type
simulations are still missing. The present version of the eikonal
cascade model${^{22,23}}$ ECCO, is less refined${^2}$ than the more 
conventional models constructed (in principle) for description
of quantities admitting only continuous variations in density 
distributions.  There is a hope that new possibility will be open if 
there will work the implementation of quantum interference into 
Monte Carlo generators by modelling the generalized Wigner 
functions, as
it is proposed in${^{24}}$. This hope is amplified by the experience 
that 
several kinds of structure can be generated also by modelling the
elementary probabilities which govern the branching processes (compare 
e.g. with${^{25}}$).

\vspace{1.5cm}
{\bf R e f e r e n c e s}
\begin{description}
\baselineskip=12pt
\item{ 1.} C.P. Singh, Phys. Reports {\bf 236}, 147 (1993).
\item{ 2.} E.A. DeWolf, I.M. Dremin and W. Kittel, preprint HEN-362, 
update
          July 1995.
\item{3.} A. Bialas and R. Peschanski, Nucl. Physics B {\bf 273}, 703
          (1986).
\item{4.} A. Bialas and R. Peschanski, Nucl. Physics B {\bf 308}, 857
          (1988).
\item{5.} R.C. Hwa, Phys. Rev. D {\bf 41}, 1456 (1990).
\item{6.} C.B. Chiu and R.C. Hwa, Phys. Rev. D {\bf 43}, 100 (1991).
\item{7.} I. Derado, R.C. Hwa, G. Jancso and N. Schmitz, Phys. Lett. B
          {\bf 283}, 151 (1992).
\item{8.} R.K. Shivpuri and V. Anand, Phys. Rev. D {\bf 50}, 287 
(1994).
\item{9.} G. Singh, A. Mukhopadhyay and P.L. Jain,  Z. Phys. A 
          {\bf 345}, 305 (1993).
\item{10.} M. Bla\v zek, Z. Phys. C {\bf 63}, 263 (1994).
\item{11.} M.I. Adamowich et al., (EMU01 Collab.), Phys. Rev. Lett.
           {\bf 65}, 412 (1990).
\item{12.} I. Derado, G. Jancso, N. Schmitz and P. Stopa, Z. Phys. C
           {\bf 47}, 23 (1990).
\item{13.} K. Fialkowski, B. Wosiek and J. Wosiek, Acta Phys. 
Polonica B
           {\bf 20}, 639 (1989).
\item{14.} B. Wosiek et al. (KLMM Collab.), In: Proc. XXV-th 
Int'l Symp. on
   Multiparticle Dynamics, Star\' a Lesn\' a, Slovakia, 1995. 
World Sci.,
   Singapore, in press.
\item{15.} K. Kadija and P. Seyboth, Z. Phys. C {\bf 61}, 465 (1994). 
\item{16.} E.K. Sarkisyan, L.K. Gelovani, G.L. Gogiberidze and G.G.
           Taran,  Phys. Lett. B {\bf 347}, 439 (1995).
\item{17.} P. Lipa and B. Buschbeck, Phys. Lett. B {\bf 223}, 
465 (1989).
\item{18.} T.C. Halsey et al., Phys. Rev. A {\bf 33}, 1141 (1986).
\item{19.} M. Bla\v zek,  {\em Multifractals in High Energy Nuclear
   Collisions by Solving an Inverse Problem.} In: {\em Quantum 
Inversion
   Theory and Applications.} Ed.: H.V. von Geramb. Springer Verlag, 
Berlin
   1994; p. 389.

\smallskip
M. Bla\v zek, {\em High Energy Nuclear Collisions and Phase 
Transitions.}           
   In: {\em Multiparticle Dynamics} 1994. Ed.:  A. Giovannini, 
S. Lupia and
   R. Ugoccioni. World Sci., Singapore, 1995; p. 189.    
\item{20.} M.I. Adamowich et al. (EMU01 Collab.), Phys. Rev. Letters 
           {\bf 65}, 412 (1990).
\item{21.} T. \accent23Akeson et al. (Helios Collab.), 
           Phys. Lett. B {\bf 252}, 303 (1990).
\item{22.} R.C. Hwa and J.C. Pan, Phys. Rev. D {\bf 45}, (1992) 106.
\item{23.} R.C. Hwa and J.C. Pan, Phys. Rev. D {\bf 46}, (1992) 2941.
\item{24.} A. Bialas and A. Krzywicki, Phys. Lett. B {\bf 354}, 
(1995) 134.
\item{25.} M. Bla\v zek, Czech. J. Phys. B {\bf 32}, (1982) 617.
\item{26.} R. Holynski et al., Phys. Rev. Lett. {\bf 62}, 733 (1989). 
\item{27.} P. Carruthers, E.M. Friedlander, C.C. Shih and R.M.
           Weiner, Phys. Lett. B {\bf 222}, 487 (1989).
\item{28.} C.B. Chiu and R.C. Hwa, Phys. Rev. D {\bf 45}, 2276 (1992).
\item{29.} R.C. Hwa and J.C. Pan, Phys. Rev. D {\bf 45}, 1476 (1992).           
\item{30.} B. Bushbeck, P. Lipa and R. Peschanski, Phys. Lett. 
B {\bf 215},
           788 (1988).
\item{31.} F. Botterweck et al. (EHS-NA 22 Collab.), 
Z. Phys. C {\bf 51},
           37 (1991).
\item{32.} S. Wang et al., Phys. Rev. D {\bf 49}, 5785 (1994).
\item{33.} R. Albrecht et al. (WA 80 Collab.), Phys. Lett. B 
{\bf 221}, 427
           (1989).
\item{34.} A. Bialas and M. Gazdzicki, Phys. Lett. B {\bf 252}, 
483 (1990).
\item{35.} E.K. Sarkisyan, L.K. Gelovani, G.G. Taran and G.I. Sakharov,
           Phys. Lett. B {\bf 318}, 568 (1993).
\end{description}
\vfill  \eject

\vspace{1.5cm}
\begin{center}
Table captions
\end{center}

\vspace{1cm}
TABLE 1.
Experimental slopes $a(q)$, $\tau (q)$ and the expression $\Delta(q)
\equiv a(q) + \tau (q) - (q-1)$ characterizing fractal structure of 
(pseudo)rapidity distributions appearing in three different types
of collisions. The quantity $\Delta (q)$ is accompanied by mean
quadratic error. The slopes of Ref.${^7}$ and Ref.${^8}$ concern the 
case
[X(A/C)] of Table 2  while those of Ref.${^9}$ remind us there 
the case [Y(A/C)].

\vspace{1cm}
TABLE 2.
Several expressions of numerator and denominator entering definition of 
the factorial moments which are investigated when searching  for 
scaling
properties of high energy density fluctuations. In some references the 
corresponding moments are involved up to an overall normalization 
factor. 
The $q$-th power factorial multinomial is defined by $F(n;q) \equiv 
n(n-1) \ldots (n-q+1)$.

\vfill  \eject

\begin{table}[t]       

\vspace{-5cm}        
\begin{center}
TABLE 1.  \\

\vspace{2cm}
\begin{tabular}{|c|}
\hline
 ~ \\
 $\mu $-$p$ and $\mu$-$d$ collisions 
        at 280 GeV, data from${^7}$     \\
\begin{tabular}{|c|c|c|c|}
\hline
   q  &   $a(q) $          &    $\tau(q)       $   & $\Delta(q)$ \\
\hline
   2  &  0.012 $\pm$ 0.002  &  0.91 $\pm$ 0.01  
&  -0.078 $\pm$ 0.010 \\
   3  &  0.034 $\pm$ 0.016  &  1.81 $\pm$ 0.02  & 
      -0.156 $\pm$ 0.026 \\
   4  &  0.12  $\pm$ 0.05   &  2.64 $\pm$ 0.05  &  -0.24  $~\pm$  
      0.07 \\
   5  &  0.22  $\pm$ 0.11   &  3.57 $\pm$ 0.12  &  -0.21  
  $~\pm$  0.16 \\
\hline
\end{tabular}
 ~ \\
  ~\\
 ~  \\
proton-nucleus int's at 800 GeV/c, data from${^8}$  \\
\begin{tabular}{|c|c|c|c|}
\hline
      q  &  $a(q)$  &     $\tau(q)$   &    $\Delta(q)$ \\
\hline
      2  & 0.180 $\pm$ 0.020  &  0.694 $\pm$ 0.005  & -0.126 
          $\pm$ 0.021 \\
      3  & 0.484 $\pm$ 0.040  &  1.269 $\pm$ 0.014  & -0.247 
          $\pm$ 0.042 \\
      4  & 0.947 $\pm$ 0.050  &  1.740 $\pm$ 0.024  & -0.313 
          $\pm$ 0.055 \\
      5  & 1.516 $\pm$ 0.072  &  2.032 $\pm$ 0.052  & -0.452 
          $\pm$ 0.089 \\
      6  & 2.035 $\pm$ 0.120  &  2.493 $\pm$ 0.067  & -0.472 
          $\pm$ 0.137 \\
\hline
\end{tabular}
 ~ \\
 ~ \\
  ~ \\
$^{84}$Kr at 1.52 A GeV colliding  
with $Em$, data from${^9}~~$ \\
\begin{tabular}{|c|c|c|c|}
\hline
     q   &    $a(q) $  &   $ \tau(q)$  &  $\Delta(q)$  \\
\hline
   2  &  0.023 $\pm$ 0.003  &  0.645 $\pm$ 0.028  & -0.332 
     $\pm$ 0.028 \\
   3  &  0.069 $\pm$ 0.008  &  1.238 $\pm$ 0.058  & -0.693 
     $\pm$ 0.058 \\
   4  &  0.117 $\pm$ 0.017  &  1.802 $\pm$ 0.090  & -1.081 
     $\pm$ 0.092 \\
   5  &  0.148 $\pm$ 0.024  &  2.347 $\pm$ 0.122  & -1.505 
     $\pm$ 0.124 \\   
\hline
\end{tabular}    
 ~  \\
 ~ \\
\hline
\end{tabular}
 ~~ \\
  ~~ \\
   ~~ \\
    ~~ \\
     ~~ \\
\end{center}
\end{table}

\begin{table}[t]
\vspace{-1cm}
\tabcolsep-.05mm
\begin{center}
TABLE 2.  \\

\vspace{1cm}
\hspace*{-1cm}
\begin{tabular}{|c|c|c|c|c|}  
\hline
\hline
& & & &  \\

\vspace{-.2cm}
   &     & {\bf X}  &  {\bf Y}   & {\bf Z}     \\ 
   &     & $~~~~ ~~~~ ~~~~ ~~~~ ~~~ ~~~~ ~~$[Ref.]  &
            $ ~~~~ ~~~~ ~~~~ ~~~~ ~~~~ ~~~~ ~~$[Ref.]$~$ &
             $~~~~ ~~~~ ~~~~ ~~~~ ~~~~ ~~~~ ~~$ [Ref.]   \\
\hline
 & &  & & \\
  & {\bf A} &  $\frac{1}{M} \sum\limits_{m=1}^M
          F(n_{me};q)$ &
       $ \frac{1}{E} \sum\limits_{e=1}^E F(n_{me};q)$ &
        $\frac{1}{ME} \sum\limits_{m=1}^M \sum\limits_{e=1}^E 
             F(n_{me};q)$ \\
  Numerator &  & & &  \\
  &  & horizontal    & vertical  &  horizontal+vertical  \\
  &  &  factorial   moments & factorial moments &  factorial 
             moments \\
\hline
 & &  & &  \\
  Notation & {\bf B} &  $\sum\limits_{m=1}^M n_{me}~=~ N_e^{(H)}$ & $
     \sum\limits_{e=1}^E n_
     {me}~=~N_m^{(V)}$ & $ \sum\limits_{m=1}^M \sum\limits_{e=1}^E n_
     {me}~=~ N^{(HV)}~~$  \\ 
 &   &    &  &   \\
\hline
  & {\bf C} & $~~\left [ \frac{1}{M} N_e^{(H)} 
        \right ]^{{{}^{q^{\phantom{X}}}}}  ~~~
        ~~~~[{\rm R}^{(H)}]~$ &
        $~~\left [ \frac{1}{E} N_m^{(V)} \right ]^{{}^q} ~~ ~ 
        ~~~~  ~~~~ [{\rm R}^{(V)}]~ $ &
        $~~~\left [ \frac{1}{EM} N^{(HV)}  \right ]^{{}^q} ~~~ 
        ~~[{\rm R}^{(HV)}] $  \\
  & & & & \\
  & $ {\bf D_1} $& -------- &  -------- &~ $~~\sum\limits_{m=1}^M 
        [N_m^{(V)}]^{{}^q }
        ~~~   ~~~~   ~~  ~~ [{\rm R}{^{12}}]~ $ \\
  Several & & & & \\
  & ${\bf D_2}$ & -------- & -------- & $~~ ~~\sum\limits_{e=1}^E 
        [N_e^{(H)}]^{{}^q}$ ~~~~ ~~~~  \\
  expressions& & & & \\
  &{\bf E} & $\left ( \frac{1}{M} \right )^{{}^q} F(N_e^{(H)};q)~~~
[{\rm R}{^{3,4}}]
$ &  $\left ( \frac{1}{E} \right ) ^{{}^q} \! F(N_m^{(V)};q) ~~
[{\rm R}{^{15,16}}] 
$& $\left ( \frac{1}{ME} \right )^{{}^q} F(N^{(HV)};q) ~~~~  ~~~~$  \\
  of & & & &  \\
 &{\bf F}& -------- & -------- & $\left ( \frac{1}{E} \right )^
      {{}^q} F(
       \frac{1}{M} N^{(HV)} ;q) ~~[{\rm R}{^{15}}] $  \\
 denominator  & & & & \\
 &  ${\bf G_1}$ &   -------- &  --------& $\frac{1}{M} 
     \sum\limits_{m=1}^M 
       \left ( \frac{1}{E} \right ) ^q F(N_m^{(V)};q)$~~  \\

\vspace{-.2cm}
  & & & &  $ ~~~  ~~ ~~~~ ~~~~ ~~~~ ~~~~ ~~[{\rm R}{^{14,15}}]$  \\
  & & & &  \\
  &  ${\bf G_2}$ & -------- & -------- & $\frac{1}{M}  
       \sum\limits_{m=1}^
    M \left( \frac{1}{M} \right )^{q-1} \! F(\frac{1}{E} 
        N_m^{(V)};q) $ \\
  &   &     &  &  \\
\hline
  Full form &  & & & \\
  involves an  & {\bf V} & vertical averaging & 
                     horizontal averaging &   -------- \\          
  overall & & & & \\
\hline
  Are the & & & & \\
 effective  & & & &  \\
 average & {\bf W}  &  yes &  yes & no  \\
 multiplicities & & & &  \\
 introduced ? & & & & \\     
\hline
\hline
\end{tabular}
\end{center}
   ~~ \\
 $~~~~$\phantom{QQ} References in line C,$~~~\! [{\rm R}^{(H)}]$:$~~$ 
             Ref.${^{3,4,7,8,11,26,27,28,29}}$;  \\
\phantom{QQReferences in line C,S~} $[{\rm R}^{(V)}]$:$~~$ 
             Ref.${^{9,13,23,27,30,31,32}}$;  \\
\phantom{QQReferences in line C,S~} $[{\rm R}^{(HV)}]: \!$ 
             Ref.${^{4,12,31,32,33,34,35}}$.  \\
 ~ \\
\end{table}
\end{document}